\documentstyle[amssymb,preprint,aps]{revtex}

\begin{document}
\title{Schr\"{o}dinger equation from an exact uncertainty principle}
\author{Michael J. W. Hall$^{1}$ and Marcel Reginatto$^{2,}$\thanks{%
Present address: Physikalisch-Technische Bundesanstalt, Bundesallee 100,
38116 Braunschweig, Federal Republic of Germany}}
\address{$^{1}${\it Theoretical Physics, IAS, Australian National University,}\\
{\it Canberra ACT 0200, Australia}\\
$^{2}${\it U. S. Department of Energy, }\\
{\it Environmental Measurements Laboratory, }\\
{\it New York, New York 10014-4811, USA}}
\maketitle
\pacs{}

\begin{abstract}
An exact uncertainty principle, formulated as the assumption that a
classical ensemble is subject to random momentum fluctuations of a strength
which is determined by and scales inversely with uncertainty in position,
leads from the classical equations of motion to the Schr\"{o}dinger equation.

PACS: 03.65.Bz
\end{abstract}
\tightenlines
\renewcommand{\thesection}{\arabic{section}}
\renewcommand{\thesubsection}{\thesection.\arabic{subsection}}
\section{Introduction}

The uncertainty principle is generally considered to be a fundamental
conceptual tool for understanding differences between classical and quantum
mechanics. As first argued by Heisenberg in 1927 \cite{Heisenberg27}, the
fact that quantum states do not admit simultaneously precise values of
conjugate observables, such as position and momentum, does not necessarily
imply an incompleteness of the theory, but rather is consistent with not
being able to simultaneously determine such observables experimentally to an
arbitrary accuracy.

Corresponding uncertainty relations such as $\Delta x\Delta p\geq \hbar /2$
``give us that measure of freedom from the limitations of classical concepts
which is necessary for a consistent description of atomic processes'' \cite
{Heisenberg30}. The uncertainty principle provides the basis of the
Copenhagen interpretation of quantum mechanics, famously used by Bohr in
defending the completeness of the theory against critics such as Einstein 
\cite{Bohr58}.

If regarded as merely asserting a physical limit on the degree to which
classical concepts can be applied, the uncertainty principle is not
sufficiently restrictive in content to supply a means for moving from
classical mechanics to quantum mechanics. Thus Landau and Lifschitz write
that ``this principle in itself does not suffice as a basis on which to
construct a new mechanics of particles'' \cite{Landau77}. In particular,
uncertainty relations expressed as imprecise inequalities are not enough to
pin down the essence of what is nonclassical about quantum mechanics.
Authors have tended to point instead, for example, to the commutation
relation $[\widehat{x},\widehat{p}]=i\hbar $ for quantum observables \cite
{Born25}, or to the principle of superposition of wavefunctions
\cite{Landau77,Dirac58}, in this regard.

However, it will be shown here that an {\it exact} form of the uncertainty
principle may in fact be formulated, which provides the {\it single} key
element in moving from the equations of motion of a classical ensemble to
those of a quantum ensemble. In particular, if it is assumed that a
classical ensemble is subject to random momentum fluctuations, {\it where
the strength of these fluctuations is precisely determined by and scales
inversely with uncertainty in position} (as characterised by the position
probability density), then the resulting modified equations of motion are
equivalent to the Schr\"{o}dinger equation. Thus, surprisingly, there is an
exact formulation of the uncertainty principle which does in fact capture
the essence of what is ``quantum'' about quantum mechanics.

In the following section we recall the description of a classical ensemble
in terms of a pair of equations in configuration space (the Hamilton-Jacobi
equation and the continuity equation), and provide the corresponding
Lagrangian from which these equations follow. In section 3 we show that
the above exact uncertainty principle leads to a modification of this
Lagrangian (essentially incorporating the kinetic energy of the random
momentum fluctuations), the form of which yields equations of motion
equivalent to the Schr\"{o}dinger equation. Further, an exact uncertainty 
{\it relation} for position and momentum uncertainties
is derived, corresponding to the exact uncertainty principle, from which
the usual Heisenberg inequality follows as a consequence.

Of course, equations of motion equivalent to the Schr\"{o}dinger equation do
not in themselves imply the full quantum formalism. Accordingly, in section
4 a Hamiltonian formulation is provided for the equations of motion, which
leads naturally to the usual wavefunction representation as corresponding to
the normal modes of the modified system. Conclusions are presented in
section 5.

\section{Classical Mechanics}

For simplicity, we limit ourselves to the case of a single particle,
described in a configuration space of $n$ dimensions. \ In the
Hamilton-Jacobi formulation of classical mechanics, the equation of motion
takes the form \cite{Corben&Stehle94} 
\begin{equation}
\frac{\partial S}{\partial t}+\frac{1}{2m}\nabla S\cdot \nabla S+V=0.
\label{CHJ}
\end{equation}
The velocity field ${\bf u}({\bf x},t)$ that describes the motion of the
particle is related to the momentum potential $S({\bf x},t)$ by 
\begin{equation}
{\bf u}=\frac{1}{m}\nabla S.  \label{dsdx}
\end{equation}
We assume that the initial conditions are not known exactly, and that the
probability of finding the particle in a given volume of the configuration
space is described by a probability density $P({\bf x},t)$. \ The
probability density must satisfy the following two conditions: it must be
normalized, 
\[
\int Pd^{n}x=1, 
\]
and it must satisfy a continuity equation, 
\begin{equation}
\frac{\partial P}{\partial t}+\nabla \cdot \left( P\frac{1}{m}\nabla
S\right) =0.  \label{CONT}
\end{equation}

Eqs. (\ref{CHJ}) and (\ref{CONT}), together with (\ref{dsdx}), completely
determine the motion of the classical ensemble. \ Eqs. (\ref{CHJ}) and (\ref
{CONT}) can be derived from the Lagrangian 
\begin{equation}
L_{C}=\int P\left\{ \frac{\partial S}{\partial t}+\frac{1}{2m}\nabla S\cdot
\nabla S+V\right\} d^{n}xdt  \label{CL}
\end{equation}
by fixed end-point variation ($\delta P=\delta S=0$ at the boundaries) with
respect to $S$ and $P$.

\section{The transition from classical mechanics to quantum mechanics}

\subsection{Momentum fluctuations}

Consider now the possibility that the classical Lagrangian is not quite
right, because $\nabla S$ is actually an average momentum: one also has a
fluctuation ${\bf N}$ about $\nabla S$. \ Thus the physical momentum is \[%
{\bf p}=\nabla S+{\bf N}.\]
No particular underlying physical model will be assumed for the momentum
fluctuation ${\bf N}$. Indeed, one could instead regard the fluctuations
as fundamentally nonanalyzable, being introduced as a simple device to
remove the notion of individual particle trajectories (since ${\bf u}$
in Eq. (\ref{dsdx}) is no longer ascertainable).

Since the momentum fluctuation ${\bf N}$ may conceivably depend on
position, the average over such fluctuations for a given
quantity $A$ at point ${\bf x}$ will be denoted by $\overline{A}$, while 
the average over fluctuations {\it and} position will be denoted by $<A>$.
One hence has the general relation
$<A> = \int P\overline{A} d^n x .$
A physically very reasonable {\it randomness} assumption for the 
momentum fluctuation ${\bf N}$ is that it vanishes on average
everywhere, i.e., $\overline{{\bf N}}\equiv 0$.  However, here only two
weaker assumptions will be made: 
\begin{equation} \label{ranassump}
<{\bf N}>=0, \,\,\,\,\,\,\,\,\,\,\,\,\,\,\,   <\nabla S\cdot {\bf N}> = 0. 
\end{equation}
The first of these states that the fluctuations are unbiased, and the
second that the fluctuations are linearly uncorrelated with the average momentum
$\nabla S$. 
 
It follows that when the momentum
fluctuations are significant, 
the kinetic energy term $<\frac{1}{2m}\nabla S\cdot \nabla S>$
in the Lagrangian should be replaced by 
by $<\frac{1}{2m}(\nabla S+{\bf N})\cdot (\nabla S+{\bf N})>$, 
yielding the modified Lagrangian 
\begin{eqnarray}
L &=&\int P\left\{ \frac{\partial S}{\partial t}+\frac{1}{2m}\overline{\left( 
\nabla S+%
{\bf N}\right) \cdot \left( \nabla S+{\bf N}\right)} +V\right\} d^{n}xdt 
\nonumber \\
&=&\int P\left\{ \frac{\partial S}{\partial t}+\frac{1}{2m}\nabla S\cdot
\nabla S+V\right\} d^{n}xdt+\frac{1}{2m}\int P \overline{{\bf N}\cdot {\bf N}}
d^{n}xdt  \nonumber \\
&\equiv &L_{C}+\frac{1}{2m}\int \left( \Delta N\right) ^{2}dt  
\label{modified lagrangian}
\end{eqnarray}
where $\Delta N$ is the average rms momentum fluctuation, given by 
$<{\bf N}\cdot {\bf N}>^{1/2}$. 
Thus the
consequence of taking into consideration the momentum fluctuations is to add
a positive term to the Lagrangian, arising from the additional kinetic
energy due to the fluctuations.  

\subsection{Exact uncertainty principle}

How can we estimate the magnitude of this additional term, if we don't know
anything else about the system except the probability density $P$ and the
average momentum $\nabla S$? To estimate the magnitude of the momentum
spread, we will assume that an exact uncertainty principle holds, in the
sense that the strength of the momentum fluctuations at a given time are
inversely correlated with uncertainty in position at that time, where the
uncertainty of position is characterized by $P$. \ Clearly, this assumption
is an additional hypothesis that is {\em independent} of classical
mechanics. \ 

To make this assumption precise, consider the general case of an $n$%
-dimensional space and a one-parameter family of probability distributions
(which we label with a parameter $k>0$) at time $t_{0}$, related by a
rescaling of variables 
\[
P({\bf x})\rightarrow P_{k}({\bf x})\equiv k^{n}P(k{\bf x}) . 
\]
These transformations preserve the normalization,

\[
\int P({\bf x})d^{n}x\rightarrow \int k^{n}P(k{\bf x})d^{n}x=\int P({\bf y}%
)d^{n}y 
\]
where we have introduced the change of variables ${\bf y}=k{\bf x}.$ \ We
also have

\begin{eqnarray*}
\nabla P({\bf x})\cdot \nabla P({\bf x}) &\rightarrow &k^{2n+2}\nabla _{y}P(%
{\bf y})\cdot \nabla _{y}P({\bf y}) \\
{\bf x}\cdot \nabla P({\bf x}) &\rightarrow &k^{n}{\bf y}\cdot \nabla _{y}P(%
{\bf y}).
\end{eqnarray*}

Under such a transformation, any direct measurement of position
uncertainty $\delta x$\ such as the rms uncertainty $\Delta x$\ changes
according to the rule

\[
\delta x\rightarrow \delta x_{k}\equiv \frac{1}{k}\left( \delta x\right) . 
\]
Thus probability densities with different values of $k$ represent physical
systems that {\em only differ in how well we know the location of the
particle}, since the shape of the probability densities are the same except
for the rescaling. The exact uncertainty principle that we want to make
use of corresponds roughly to the assumption that such a 
scaling of position by a factor $1/k$ scales the momentum fluctuation by a
factor $k$. 

More precisely, the exact uncertainty principle is 
equivalent to the statement that {\it the momentum fluctuation }$\Delta N$%
{\it \ is determined by the uncertainty in position, where the latter is
characterised by the probability density }$P${\it , and where} 
\begin{equation}
\Delta N\rightarrow k\Delta N{\it \ }  \label{deltaN}
\end{equation}
{\it \ under }$k${\it \ transformations}.  Note that the uncertainty
product $\delta x\Delta N$ is thus preserved under $k$ transformations,
for any direct measure of position uncertainty $\delta x$.

To apply the exact uncertainty principle, 
we assume that the Lagrangian formalism remains 
applicable to $L$.  Hence the additional term in Eq. (\ref{modified
lagrangian}) must be a spacetime integral over a
scalar function of ${\bf x}$, $P$ and $S$ and their derivatives. \ Moreover,
since $\left( \Delta N\right) ^{2}$ is determined solely by position
uncertainty (where the latter is characterised by $P$), then this
additional term is in fact independent of $S$. \ Finally, for causality to
be preserved (i.e., the equations of motion require only $P$ and $S$ to be
specified on an initial surface), second and higher order derivatives of $P$
must be excluded\footnote{%
Requirements of causality do not exclude a term linear in $\nabla ^{2}P$,
but since it can be shown that such a term does not lead to a different
result we will not consider it here.}. \ Hence, the additional term in the
Lagrangian (\ref{modified lagrangian}) can be written in the form

\begin{equation}
\int \left( \Delta N\right) ^{2}dt=\int Pf({\bf x},P,{\bf x}\cdot \nabla
P,\nabla P\cdot \nabla P)d^{n}xdt.  \label{intPf}
\end{equation}

The exact uncertainty principle requires $f$ to transform under $k$
transformations as follows, 
\begin{eqnarray*}
&&\int P({\bf x})f\left[ {\bf x},P({\bf x}),{\bf x}\cdot \nabla P({\bf x}%
),\nabla P\cdot \nabla P({\bf x})\right] d^{n}xdt \\
&\rightarrow &\int P({\bf y})f(k^{-1}{\bf y},k^{n}P({\bf y}),k^{n}{\bf y}%
\cdot \nabla _{y}P({\bf y}),k^{2n+2}\nabla _{y}P({\bf y})\cdot \nabla _{y}P(%
{\bf y}))d^{n}ydt \\
&\equiv &k^{2}\int P({\bf y})f({\bf y},P({\bf y}),{\bf y}\cdot \nabla _{y}P(%
{\bf y}),\nabla _{y}P({\bf y})\cdot \nabla _{y}P({\bf y}))d^{n}ydt.
\end{eqnarray*}
This leads to the homogeneity condition

\begin{equation}
f(k^{-1}{\bf x},k^{n}u,k^{n}v,k^{2n+2}w)=k^{2}f({\bf x},u,v,w), 
\label{homogeneity}
\end{equation}
where we have introduced the more compact notation 
\begin{eqnarray}
u &=&P  \nonumber \\
v &=&{\bf x}\cdot \nabla P  \label{uvw} \\
w &=&\nabla P\cdot \nabla P . \nonumber
\end{eqnarray}
From this requirement we derive the first order partial differential equation

\begin{equation}
-\sum_{i=1}^{n}x_{i}\frac{\partial f}{\partial x_{i}}+nu\frac{\partial f}{%
\partial u}+nv\frac{\partial f}{\partial v}+(2n+2)w\frac{\partial f}{%
\partial w}=2f.  \label{1storderpde}
\end{equation}
The problem of finding the general integral of such an equation is
equivalent to the problem of finding the general integral of a system of
ordinary differential equations \cite{Schuh68}, which in our case is given by

\[
-\frac{dx_{1}}{x_{1}}=...=-\frac{dx_{n}}{x_{n}}=\frac{du}{nu}=\frac{dv}{nv}=%
\frac{dw}{(2n+2)w}=\frac{df}{2f}. 
\]
This system of ordinary differential equations has ($n$+3) independent
integrals, which can be chosen as

\begin{eqnarray*}
u^{-1}w^{1/2}x_{i} &=&const. \\
u^{-1}v &=&const. \\
u^{2/n}{\bf x}\cdot {\bf x} &=&const. \\
u^{2}w^{-1}f &=&const.
\end{eqnarray*}
and the solution of the first order partial differential Eq. (\ref
{1storderpde}) is then of the general form 
\begin{equation}
f=\left( u^{-2}w\right) g\left( u^{-1}w^{1/2}{\bf x},u^{-1}v,u^{2/n}{\bf x}%
\cdot {\bf x}\right)  \label{generalformf}
\end{equation}
where $g$ is an arbitrary function.

\subsection{Independent subsystems}

To determine $f$ completely, we need to fix the form of $g$ in Eq. 
(\ref{generalformf}). \ We therefore
introduce a natural {\it independence} condition, by the
requirement that the 
Lagrangian $L$ decomposes into additive
subsystem contributions for the case of a system composed of
independent subsystems. \ This is equivalent to the condition that the
momentum fluctuations $N_{1}$ and $N_{2}$ are linearly uncorrelated for two
such subsystems, and hence can equivalently be interpreted as a further
randomness assumption for the momentum fluctuations.

To investigate the requirements imposed on $f$ by the independence
condition, it will be sufficient
to consider the case where we have a system consisting of two uncorrelated
particles of mass $m$ that do not interact, one particle described by a set
of coordinates ${\bf x}_{1}$ and the other by ${\bf x}_{2}$. \ Thus $P$ is
of the form 
\begin{equation}
P({\bf x}_{1},{\bf x}_{2})=P_{1}({\bf x}_{1})P_{2}({\bf x}_{2}) ,
\label{pequalpip}
\end{equation}
and hence from Eqs. (\ref{modified lagrangian}) and (\ref{intPf}) the
independence condition requires  
\begin{equation}
Pf=P_{1}P_{2}\left( f_{1}+f_{2}\right)  \label{systemindep}
\end{equation}
where subscripts 1 and 2 refer to subsystems 1 and 2 respectively.

Eq. (\ref{pequalpip}) immediately implies the relations
\begin{eqnarray*}
u &=&u_{1}u_{2} \\
v^\prime &\equiv &u^{-1}v=u_{1}^{-1}v_{1}+u_{2}^{-1}v_{2}=v_1^\prime +v_2^\prime \\
w^\prime &\equiv &u^{-2}w\ =u_{1}^{-2}w_{1}+u_{2}^{-2}w_{2}=
w_1^\prime +w_2^\prime ,
\end{eqnarray*}
and hence Eq. (\ref{generalformf}) becomes 
\[
f=\left( w^\prime _{1}+w^\prime _{2}\right) g\left( \sqrt{w^\prime _{1}+w^\prime
_{2}}{\bf x},\ v^\prime _{1}+v^\prime _{2},\ \left( u_{1}u_{2}\right) ^{2/n}%
{\bf x}\cdot {\bf x}\right) , 
\]
where ${\bf x}=({\bf x}_{1},{\bf x}_{2})$. \ 
From Eq. (\ref{systemindep}), this 
form of $f$ must decompose into the sum of a function of $u_{1}$, $%
v^\prime _{1}$, $w^\prime _{1}$and ${\bf x}_{1}$, and a function of $u_{2}$, $%
v^\prime _{2}$, $w^\prime _{2}$ and ${\bf x}_{2}$. Since the factor that
multiplies $g$, and the second and third arguments of $g$, are such functions
(with respect to $w^\prime $,\ $v^\prime $, and ${\bf x}$ respectively), these
terms cannot be mixed by the functional form of $g$. \ 
It follows that $g$ must be of the form 
\[
g\left( {\bf a},b,c\right) =C+g_{0}({\bf a})+bg_{1}({\bf a})+cg_{2}({\bf a})
\]
where $C$ is a constant, and the functions $g_{j}$ satisfy the
condition 
\begin{equation}
g_{j}(\lambda {\bf a})=\lambda ^{-2}g_{j}({\bf a})\ ,\ j=0,1,2
\label{homogeneitycondg}
\end{equation}
(to allow cancellation of the factor $w^\prime _{1}+w^\prime _{2}$ that
multiplies $g$). Hence 
\[
f=C\left( w^\prime _{1}+w^\prime _{2}\right) +g_{0}({\bf x}_{1},{\bf x}%
_{2})+\left( v^\prime _{1}+v^\prime _{2}\right) g_{1}({\bf x}_{1},{\bf x}%
_{2})+\left( u_{1}u_{2}\right) ^{2/n}\left( {\bf x}_{1}\cdot {\bf x}_{1}{\bf %
+x}_{2}\cdot {\bf x}_{2}\right) g_{2}({\bf x}_{1},{\bf x}_{2}) . 
\]

The independence condition Eq. (\ref{systemindep}) places further
conditions
on the $g_{j}$. First, $g_{0}$ is required to be a sum of a function of $%
{\bf x}_{1}$ and a function of ${\bf x}_{2}$. Hence it only represents a
classical additive potential term (that satisfies
the homogeneity condition (\ref
{homogeneitycondg}) above), and so can be ignored as having no nonclassical
role (it can be absorbed into the classical potential $V$ in the
Lagrangian). Second, to avoid subsystem cross terms, $g_{1}$ must be
constant. \ But the homogeneity condition (\ref{homogeneitycondg}) is
then only satisfied by the choice $g_{1}=0$. Third, cross terms in $u_{1}$ and
$u_{2}$ can only be avoided by choosing $g_{2}=0$. 

The form of $f$ thus reduces to the first term,  
$C\left( w^\prime _{1}+w^\prime _{2}\right)$. From Eq. (\ref
{systemindep}) this term is to be identified with the sum of $f_{1}$ and $f_{2}$, thus
yielding the final form 
\begin{equation}
f=Cw^\prime =C\frac{1}{P^{2}}\nabla P\cdot \nabla P  \label{fequalFI}
\end{equation}
for $f$, where $C$ is a universal constant. Note from Eq.(\ref{intPf})
that $C$ must be positive. 

\subsection{Equations of motion}

The modified Lagrangian follows from Eqs. (\ref{modified lagrangian}), (\ref
{intPf}) and (\ref{fequalFI}) as 
\begin{equation}
L=\int P\left\{ \frac{\partial S}{\partial t}+\frac{1}{2m}\nabla S\cdot
\nabla S+\frac{C}{2m}\frac{1}{P^{2}}\nabla P\cdot \nabla P+V\right\}
d^{n}xdt.  \label{Qlagrang}
\end{equation}
Fixed end-point variation with respect to $S$ leads again to (\ref{CONT}),
while fixed end-point variation with respect to $P$ leads to 
\begin{equation}
\frac{\partial S}{\partial t}+\frac{1}{2m}\nabla S\cdot \nabla S+\frac{C}{2m}%
\left[ \frac{1}{P^{2}}\nabla P\cdot \nabla P-\frac{2}{P}\nabla ^{2}P\right]
+V=0 . \label{QHJEQ}
\end{equation}
Eqs. (\ref{CONT}) and (\ref{QHJEQ}) are identical to the Schr\"{o}dinger
equation 
\[
i\hbar \partial\psi /\partial t = -(\hbar^2/2m)\nabla^2\psi+V\psi ,\]
provided the wave function $\psi ({\bf x},t)$ is written in terms
of $S$ and $P$ by 
\[
\psi =\sqrt{P}\exp (i\frac{S}{\hbar }) 
\]
and the constant $C$ is set equal to 
\[
C=\left( \frac{\hbar }{2}\right) ^{2}. 
\]
Just {\it why} one would introduce the wavefunction $\psi $ at all is
considered in section 4 below.

Note that the classical limit of the Schr\"{o}dinger theory is not the
Hamilton-Jacobi equation for a classical particle, but Eqs. (\ref{CHJ}) and (%
\ref{CONT}) which describe a classical ensemble.

\subsection{Exact uncertainty relation}

The Schr\"{o}dinger equation has been derived above using an exact
uncertainty principle to fix the strength of random momentum fluctuation in
terms of the uncertainty in position. Note that no specific measure of
position uncertainty was assumed; it was required only that the momentum
fluctuations scale inversely with position uncertainty under $k$
transformations. \ However, having obtained a unique form, Eq. (\ref
{fequalFI}), for the function $f$ in (\ref{intPf}) we are now in a position
to write down an exact uncertainty {\it relation} relating position and
momentum uncertainties. 

For simplicity we consider the case of one dimension
($n=1$), and define 
\[
\delta x=\left[ \int P\left( \frac{1}{P}\frac{dP}{dx}\right) ^{2}dx\right]
^{-1/2}.
\]
For the case of a Gaussian probability density with rms uncertainty $\sigma $
one has $\delta x=\sigma $. \ More generally, this measure has units of
position, scales appropriately with $x$ ($\delta y=\lambda \delta x$ for $%
y=\lambda x$), and vanishes in the limit that $P$ approaches a delta
function. Hence it represents a direct measure of uncertainty for position.
From Eqs.(\ref{intPf}) and (\ref{fequalFI}) one has 
\begin{equation}
\delta x\Delta N=\sqrt{C}=\frac{\hbar }{2} . \label{eurel}
\end{equation}
Thus we have an {\it exact} uncertainty relation between position and momentum.
A quantum-operator form of this relation has been derived elsewhere,
in which $\Delta N$ is replaced by the rms deviation of a
nonclassical momentum operator \cite{Hall00}. \ 

The usual Heisenberg uncertainty relation can be derived from the above exact
uncertainty relation. From the Cramer-Rao inequality of statistical
estimation theory \cite{Cox74} one has $\Delta x\geq \delta x$, while the 
randomness assumptions in
Eqs. (\ref{ranassump}) imply 
\[
\left( \Delta p\right) ^{2}={\rm Var}(dS/dx+N)={\rm Var}(dS/dx)+\left( \Delta
N\right) ^{2}\geq \left( \Delta N\right) ^{2}, 
\]
and hence it follows immediately from Eq. (\ref{eurel}) that $\Delta x\Delta p\geq \hbar /2$.

\section{Hamiltonian formulation and wave function representation}

\subsection{Hamiltonian formulation}

In the previous section, we derived an extension of the classical Lagrangian
which yields equations of motion equivalent to the Schr\"{o}dinger equation.
\ The Lagrangian field formalism was conveniently used because it is well
known. However, one can in fact obtain equivalent results using the {\it %
Hamiltonian} form of field theory, with no essential differences in the
assumptions and manipulations used. \ 

The Hamiltonian formalism does provide one important advantage: the concept
of canonical transformations. In the previous section, the wavefunction
representation $\psi =\sqrt{P}\exp (i\frac{S}{\hbar })$ was simply
``magicked out of thin air'', to obtain the Schr\"{o}dinger equation written
in terms of the wavefunction $\psi $ instead of the hydrodynamical variables 
$P$ and $S$. \ In contrast, in the Hamiltonian formalism this complex
combination of $P$ and $S$ arises immediately from asking a natural question
about canonical transformations. \ 

The Hamiltonian form corresponding to Lagrangian (\ref{Qlagrang}) is given by

\begin{equation}
H=\int P\left\{ \frac{1}{2m}\nabla S\cdot \nabla S+\frac{\hbar ^{2}}{8m}%
\frac{1}{P^{2}}\nabla P\cdot \nabla P+V\right\} d^{n}x\equiv \int {\cal H}%
d^{n}x.  \label{qhamiltonian}
\end{equation}
The field $P$ plays the role of a field coordinate, and $S$ the role of the
momentum canonically conjugate to $P$. \ The equations of motion are given
by \cite{Corben&Stehle94}

\begin{eqnarray*}
\frac{\partial P}{\partial t} &=&\left\{ P,H\right\} =\frac{\delta H}{\delta
S} \\
\frac{\partial S}{\partial t} &=&\left\{ S,H\right\} =-\frac{\delta H}{%
\delta P}
\end{eqnarray*}
where the Poisson bracket of two functions $F$ and $G$ is defined by

\begin{equation}
\left\{ F(P({\bf x}),S({\bf x})),G(P({\bf x}^{\prime }),S({\bf x}^{\prime
}))\right\} =\int \left[ \frac{\delta F({\bf x})}{\delta P({\bf x}^{\prime
\prime })}\frac{\delta G({\bf x}^{\prime })}{\delta S({\bf x}^{\prime \prime
})}-\frac{\delta F({\bf x})}{\delta S({\bf x}^{\prime \prime })}\frac{\delta
G({\bf x}^{\prime })}{\delta P({\bf x}^{\prime \prime })}\right]
d^{n}x^{\prime \prime } . \label{pbs}
\end{equation}
To simplify the formulae, we will sometimes use the notation $P\equiv P({\bf %
x})$, $P^{\prime }\equiv P({\bf x}^{\prime })$, etc. which allows us to
write Eq. (\ref{pbs}) in the concise form

\[
\left\{ F,G^{\prime }\right\} =\int \left[ \frac{\delta F}{\delta P^{\prime
\prime }}\frac{\delta G^{\prime }}{\delta S^{\prime \prime }}-\frac{\delta F%
}{\delta S^{\prime \prime }}\frac{\delta G^{\prime }}{\delta P^{\prime
\prime }}\right] d^{n}x^{\prime \prime }. 
\]
From

\begin{equation}
\frac{\delta P}{\delta P^{\prime }}=\frac{\delta S}{\delta S^{\prime }}%
=\delta ^{n}({\bf x}-{\bf x}^{\prime })  \label{poissondelta}
\end{equation}
we derive the Poisson bracket of the canonically conjugate fields,

\[
\left\{ P,S^{\prime }\right\} =\delta ^{n}({\bf x}-{\bf x}^{\prime }). 
\]
The equations of motion that correspond to $H$ are

\begin{eqnarray*}
\frac{\partial P}{\partial t} &=&\frac{\delta H}{\delta S}=-\nabla \cdot
\left( P\frac{1}{m}\nabla S\right) \\
\frac{\partial S}{\partial t} &=&-\frac{\delta H}{\delta P}=-\left[ \frac{1}{%
2m}\nabla S\cdot \nabla S+\frac{\hbar ^{2}}{8m}\left( \frac{1}{P^{2}}\nabla
P\cdot \nabla P-\frac{2}{P}\nabla ^{2}P\right) +V\right] .
\end{eqnarray*}
These equations are of course identical to (\ref{CONT}) and (\ref{QHJEQ})
which were derived using the Lagrangian formalism.

\subsection{Wavefunctions and normal modes}

The Hamiltonian form $H$ has been expressed in (\ref{qhamiltonian}) in terms
of fields which represent important physical quantities: $P$ has the
physical interpretation of a position probability density, and $S$ that of
an average momentum potential. \ However, $H$ can be rewritten in terms of
any pair of fields $\phi $ and $\chi $ without changing the physical content
provided they are related to $P$ and $S$ by a canonical transformation, 
\begin{equation}
\left\{ P,S^{\prime }\right\} =\int \left[ \frac{\delta P}{\delta \phi
^{\prime \prime }}\frac{\delta S^{\prime }}{\delta \chi ^{\prime \prime }}-%
\frac{\delta P}{\delta \chi ^{\prime \prime }}\frac{\delta S^{\prime }}{%
\delta \phi ^{\prime \prime }}\right] d^{n}x^{\prime \prime }=\left\{ \phi
,\chi ^{\prime }\right\} .  \label{cantransf}
\end{equation}
Of course, such a transformation is generally only of interest if the new
fields have some particular physical significance.

One transformation of obvious physical interest, when it exists, is to two
fields $\phi $ and $\chi $ which have uncoupled equations of motion. Such
fields label two independent physical degrees of freedom in the system, and
hence have fundamental physical significance as the ``normal modes'' of the
system. It is therefore natural to ask whether such a transformation exists
for $H$, i.e., whether there is a one-one mapping 
\begin{eqnarray*}
P &=&P(\phi ,\chi ) \\
S &=&S(\phi ,\chi )
\end{eqnarray*}
such that the fields $\phi $ and $\chi $ are uncoupled. \ It will be seen
that this question is sufficient to {\it derive} the wavefunction
representation $\psi =\sqrt{P}\exp (i\frac{S}{\hbar })$ and its complex
conjugate from the Hamiltonian $H$, as corresponding to the physical fields
describing the ``normal modes'' of the system.\ \ 

To examine the question of whether there is a canonical transformation that
will lead to uncoupled equations of motion for $\phi $ and $\chi $ we first
need to establish the following Lemma.

{\it Lemma}: A necessary condition for two conjugate fields $\phi $ and $%
\chi $ to be uncoupled is that the corresponding Hamiltonian density ${\cal H%
}^{\prime }$ has the form 
\[
{\cal H}^{\prime }=F({\bf x},\phi ,\chi )+A_{k}({\bf x},\phi ,\chi )\partial
_{k}\phi +B_{k}({\bf x},\phi ,\chi )\partial _{k}\chi +G_{jk}({\bf x},\phi
,\chi )\left( \partial _{j}\phi \right) \left( \partial _{k}\chi \right) 
\]
where $k=1,...,n$, repeated indices are summed over, and $\partial _{k}$\
denotes the partial derivative with respect to $x_{k}$. \ Furthermore, the
symmetric part of $G_{jk}$ is independent of $\phi $ and $\chi $, i.e., 
\[
G_{jk}({\bf x},\phi ,\chi )+G_{kj}({\bf x},\phi ,\chi )=2G_{jk}({\bf x}) 
\]
where $G_{jk}({\bf x})$ is symmetric with respect to $j$ and $k$.

{\it Proof}: For a Hamiltonian 
\[
H^{\prime }=\int {\cal H}^{\prime }({\bf x},\phi ,\chi ,\partial _{k}\phi
,\partial _{k}\chi )d^{n}x 
\]
the equations of motion are given by 
\begin{eqnarray*}
\frac{\partial \phi }{\partial t} &=&\frac{\partial {\cal H}^{\prime }}{%
\partial \chi }-\frac{\partial ^{2}{\cal H}^{\prime }}{\partial x_{k}\left(
\partial _{k}\chi \right) }-\frac{\partial ^{2}{\cal H}^{\prime }}{\left(
\partial \phi \right) \partial \left( \partial _{k}\chi \right) }\partial
_{k}\phi -\frac{\partial ^{2}{\cal H}^{\prime }}{\left( \partial \chi
\right) \partial \left( \partial _{k}\chi \right) }\partial _{k}\chi \\
&&-\frac{\partial ^{2}{\cal H}^{\prime }}{\partial \left( \partial _{l}\phi
\right) \partial \left( \partial _{k}\chi \right) }\partial _{kl}\phi -\frac{%
\partial ^{2}{\cal H}^{\prime }}{\partial \left( \partial _{l}\chi \right)
\partial \left( \partial _{k}\chi \right) }\partial _{kl}\chi .
\end{eqnarray*}
A similar expression is obtained for $\frac{\partial \chi }{\partial t}$. \
Since we assume that $\phi $ evolves independently of $\chi $, then in
particular no second derivatives of $\chi $ can appear in the above equation
of motion for $\phi $, and similarly, no second derivatives of $\phi $ can
appear in the corresponding equation of motion for $\chi $. \ Hence ${\cal H}%
^{\prime }$ must be linear in both $\partial _{k}\phi $ and $\partial
_{k}\chi $, and so ${\cal H}^{\prime }$ has the general form given in the
statement of the Lemma. \ Substituting this form into the equation of motion
for $\phi $ gives 
\[
\frac{\partial \phi }{\partial t}=\frac{\partial F}{\partial \chi }-\partial
_{k}B_{k}+\left( \frac{\partial A_{k}}{\partial \chi }-\frac{\partial B_{k}}{%
\partial \phi }-\partial _{j}G_{kj}\right) \partial _{k}\phi -\frac{\partial
G_{jk}}{\partial \phi }\left( \partial _{j}\phi \right) \left( \partial
_{k}\phi \right) -G_{jk}\partial _{kl}\phi 
\]
and a similar equation for $\frac{\partial \chi }{\partial t}$. \ Hence,
since $G_{jk}$ is the coefficient of $\partial _{kl}\phi $ and $\partial
_{kl}\chi $ in the respective equations of motion, the fields are uncoupled
only if the symmetric part of $G_{jk}$ is independent of both $\phi $ and $%
\chi$. $\blacksquare$

If we now express the Hamiltonian density ${\cal H}$ (\ref{qhamiltonian}) in
terms of the new variables, we find 
\begin{eqnarray*}
{\cal H} &=&P\left\{ \frac{1}{2m}\left[ \left( \frac{\partial S}{\partial
\phi }\right) ^{2}+\left( \frac{\hbar }{2}\right) ^{2}\left( \frac{\partial
\ln P}{\partial \phi }\right) ^{2}\right] \nabla \phi \cdot \nabla \phi
\right.  \\
&&\left. +\frac{1}{2m}\left[ \left( \frac{\partial S}{\partial \chi }\right)
^{2}+\left( \frac{\hbar }{2}\right) ^{2}\left( \frac{\partial \ln P}{%
\partial \chi }\right) ^{2}\right] \nabla \chi \cdot \nabla \chi \right.  \\
&&\left. +\frac{1}{m}\left[ \frac{\partial S}{\partial \phi }\frac{\partial S%
}{\partial \chi }+\left( \frac{\hbar }{2}\right) ^{2}\frac{\partial \ln P}{%
\partial \phi }\frac{\partial \ln P}{\partial \chi }\right] \nabla \phi
\cdot \nabla \chi +V\right\} .
\end{eqnarray*}
It follows immediately from the Lemma that the first two terms of ${\cal H}$
must vanish, implying 
\begin{eqnarray}
\frac{\partial S}{\partial \phi } &=&i\alpha \frac{\hbar }{2}\frac{\partial
\ln P}{\partial \phi }  \label{dsdphi} \\
\frac{\partial S}{\partial \chi } &=&i\beta \frac{\hbar }{2}\frac{\partial
\ln P}{\partial \chi }  \nonumber
\end{eqnarray}
where $\alpha ,\beta =\pm 1$. \ Since $P$ and $S$ are real, the normal modes
are therefore complex fields. Substituting Eqs. (\ref{dsdphi}) 
into the third term of ${\cal H}$%
, the Lemma further implies that 
\[
G_{jk}({\bf x})=P\left( \frac{\hbar }{2}\right) ^{2}\left( \frac{1-\alpha
\beta }{m}\right) \frac{\partial \ln P}{\partial \phi }\frac{\partial \ln P}{%
\partial \chi }\delta _{jk}=G\delta _{jk}
\]
where $G$ is a constant (the last equality follows since the second term
has no explicit ${\bf x}$ dependence). \ Now, if $\alpha =\beta $, the
Hamiltonian density reduces to ${\cal H}=P(\phi ,\chi )V$, and the inverse
transformation to the fields $P$ and $S$ then yields a Hamiltonian density
proportional to $V$, which is inconsistent with the form of ${\cal H}$. \
Therefore 
\begin{equation}
\alpha = -\beta , \hspace{1.5cm}
2P\left( \frac{\hbar }{2}\right) ^{2}\frac{1}{m}\left[ \frac{\partial \ln P}{%
\partial \phi }\frac{\partial \ln P}{\partial \chi }\right] =G.  \label{G}
\end{equation}

From Eqs. (\ref{dsdphi}) and (\ref{G}) the Hamiltonian
density ${\cal H}$ in terms of $\phi $\ and $\chi $ is
\[
{\cal H}=G\nabla \phi \cdot \nabla \chi +P(\phi ,\chi )V
\]
and the equations of motion take the simple form 
\begin{eqnarray*}
\frac{\partial \phi }{\partial t} &=&\frac{\partial P}{\partial \chi }%
V-G\nabla ^{2}\phi  \\
-\frac{\partial \chi }{\partial t} &=&\frac{\partial P}{\partial \phi }%
V-G\nabla ^{2}\chi . 
\end{eqnarray*}
For these equations to be uncoupled $P$ must be of the form 
$P=W+X\phi +Y\chi +Z\phi \chi$, which when substituted into (\ref{G})
yields   
\begin{equation}
P=\left( \frac{2}{\hbar }\right) ^{2}\frac{m}{2}G\left( \phi +K\right)
\left( \chi +L\right) ,   \label{Pandphiandchi2}
\end{equation}
where $K$ and $L$ are constants (related to $X$, $Y$ and $Z$). \ The general
form of $S(\phi ,\chi )$ is found by substitution of (\ref
{Pandphiandchi2}) in (\ref{dsdphi}) with $\alpha =-\beta $, which leads to a
pair of differential equations with solution 
\begin{equation}
S=b+i\alpha \frac{\hbar }{2}\ln \frac{\phi +K}{\chi +L},  \label{formofS}
\end{equation}
where $b$ is an arbitrary complex constant. \ 

Eqs. (\ref{Pandphiandchi2}) and (\ref{formofS}) establish the functional forms 
of $P(\phi ,\chi )$\
and $S(\phi ,\chi )$ that will permit uncoupled equations of motion for $%
\phi $ and $\chi $. \ We now have to check that these functional forms lead
to a canonical transformation. \ This requires 
\begin{eqnarray*}
\left\{ P,S^{\prime }\right\} &=&\int \left[ \frac{\delta P}{\delta \phi
^{\prime \prime }}\frac{\delta S^{\prime }}{\delta \chi ^{\prime \prime }}-%
\frac{\delta P}{\delta \chi ^{\prime \prime }}\frac{\delta S^{\prime }}{%
\delta \phi ^{\prime \prime }}\right] d^{n}x^{\prime \prime } \\
&=&-mG\left( i\alpha \frac{2}{\hbar }\right) \delta ^{n}({\bf x}-{\bf x}%
^{\prime }) \\
&=&\delta ^{n}({\bf x}-{\bf x}^{\prime })
\end{eqnarray*}
and thus the value of $G$ is fixed to be 
\[ G = i\alpha\hbar /(2m) .\] 
Recalling that $P$ and $S$ are real (and using the property that $P$ is
positive), one can show that the inverse transformation follows from (\ref
{Pandphiandchi2}) and (\ref{formofS}) as 
\begin{eqnarray}
\phi &=&a\sqrt{P}\exp \left( -i\alpha S/\hbar \right) -K  \label{phiandchi}
\\
\chi &=&\frac{\alpha \hbar }{ia}\sqrt{P}\exp \left( i\alpha S/\hbar \right)
-L  \label{chi}
\end{eqnarray}
where $a$ is an arbitrary complex constant (related to $b$).

Thus, an essentially unique canonical transformation to uncoupled fields 
$\phi $ and $\chi $ indeed exists, given by Eqs. (\ref{phiandchi}) and 
(\ref{chi}). We recognise that these fields
are, up to a scale factor and additive constant, the usual wavefunction $%
\psi =\sqrt{P}\exp (i\frac{S}{\hbar })$ and its complex conjugate, and hence
the wavefunction has a fundamental physical significance as a ``normal
mode'' of the system. The corresponding Hamiltonian density ${\cal H}$
follows as 
\begin{eqnarray*}
{\cal H} &=&\frac{i\alpha\hbar}{2m}\nabla \phi \cdot \nabla \chi
+\frac{i\alpha}{\hbar }V\left( \phi +K\right) \left( \chi +L\right) \\
&=&\frac{\hbar ^{2}}{2m}\left| \nabla \psi \right| ^{2}+V\left| \psi \right|
^{2}
\end{eqnarray*}
for all choices of $a$, $K$ and $L$, and leads directly to the
Schr\"{o}dinger equation and its conjugate.

We point out a quantization condition that follows from Eqs. (\ref{formofS}),
(\ref{phiandchi}), and (\ref{chi}). \ 
If $\Phi ({\bf x})$ is a single-valued complex
function, $\ln \Phi $ is a multi-valued function which satisfies $%
\oint_{C}d\ln \Phi =\pm i2\pi n$, where $n$ is an integer. \ If we make the
assumption that the fields $\phi $ and $\chi $ describing the ``normal
modes'' of the system are single-valued functions, then 
\[
\oint_{C}dS=-i\alpha \frac{\hbar }{2}\oint_{C}d\left( \ln \psi -\ln \psi
^{\ast }\right) =\pm 2\pi \hbar n. 
\]
This is precisely the quantization condition that was introduced by
Takabayasi\ as ``a new postulate'' in his hydrodynamic interpretation of
quantum mechanics \cite{Takabayasi52}. \ This subsidiary condition is of
course compatible with the equations of motion.

\subsection{Expectation values}

While the wavefunction and the corresponding wave equation have been
obtained, these do not represent the full quantum formalism. For example,
the nature of the assumptions about momentum fluctuations in section 3.1
provide recipes for calculating the first two moments of the momentum
distribution in terms of integrals that can now be expressed in terms of the
wavefunction, since 
\begin{eqnarray*}
&<&{\bf p}>=\int P\nabla Sd^{n}x=\frac{\hbar }{i}\int \psi ^{\ast }\nabla
\psi\, d^{n}x , \\
&<&p^{2}>=\int P\left( \nabla S\cdot \nabla S+\left( \frac{\hbar }{2}\right)
^{2}\frac{1}{P^{2}}\nabla P\cdot \nabla P\right) d^{n}x=\hbar ^{2}\int
\left| \nabla \psi \right| ^{2}d^{n}x.
\end{eqnarray*}
However, it is not immediately clear how within this framework higher-order
moments are to be calculated, nor expectation values of functions of
position and momentum. \ We briefly note here a possible approach to this
problem, based on a symmetry in the representation of position and momentum
displacements, which leads to the usual relations assumed in the Hilbert
space formulation of quantum mechanics. \ 

Under a position displacement $T_{a}:{\bf x}\rightarrow {\bf x}+{\bf a}$,
the fields $P$ and $S$ transform as $P({\bf x})\rightarrow P({\bf x}-{\bf a}%
) $, $S({\bf x})\rightarrow S({\bf x}-{\bf a})$. \ Hence in the wavefunction
representation one has 
\begin{equation}
T_{a}:\psi ({\bf x})\rightarrow \psi ({\bf x}-{\bf a}).  \label{TA}
\end{equation}
Under a momentum displacement $M_{q}:{\bf p}\rightarrow {\bf p}+{\bf q}$,
the position distribution (which is given by the field $P$) should be
unaffected, while the average momentum must change by ${\bf q}$, $\nabla
S\rightarrow \nabla S+{\bf q}$. \ Therefore, the fields $P$ and $S$
transform as $P({\bf x})\rightarrow P({\bf x})$, $S({\bf x})\rightarrow S(%
{\bf x})+{\bf q}\cdot {\bf x}$ (where an arbitrary additive constant added
to $S$ has been ignored, as it has no effect on the equations of motion). \
Hence in the wavefunction representation one has 
\begin{equation}
M_{q}:\psi ({\bf x})\rightarrow \exp (i{\bf q}\cdot {\bf x}/\hbar )\psi (%
{\bf x}).  \label{MA}
\end{equation}

Comparing (\ref{TA}) and (\ref{MA}), one recognises that the transformations 
$T_{a}$ and $M_{q}$ are Fourier-pairs. \ In particular, if one defines the
Fourier transform of $\psi ({\bf x})$ by 
\[
\varphi ({\bf p})=\frac{1}{\left( 2\pi \sigma \right) ^{-n/2}}\int \psi (%
{\bf x})\exp (i{\bf x}\cdot {\bf p}/\sigma )d^{n}x
\]
where $\sigma $ is a constant with units of action, then one has 
\begin{eqnarray*}
T_{a} &:&\varphi ({\bf p})\rightarrow \exp (i{\bf a}\cdot {\bf p}/\sigma
)\varphi ({\bf p}) \\
M_{q} &:&\varphi ({\bf p})\rightarrow \varphi ({\bf p}-{\bf \sigma q/}\hbar ) . 
\end{eqnarray*}
Comparing with Eqs. (\ref{TA}) and (\ref{MA}), there is a direct symmetry
between $\psi $ and $\varphi $ under position and momentum translations,
provided one sets $\sigma =\hbar $.

In light of this symmetry, it is natural to postulate that, in analogy to $P(%
{\bf x})=|\psi ({\bf x})|^{2}$, the momentum probability density is given by 
$\widetilde{P}({\bf p})=|\varphi ({\bf p})|^{2}$. \ Under this postulate one
finds that 
\[
<f({\bf p})>=\int \widetilde{P}({\bf p})f({\bf p})d^{n}p=\int \psi ^{\ast }(%
{\bf x})f(\frac{\hbar }{i}\nabla )\psi ({\bf x})d^{n}x , 
\]
which then leads to the natural generalisation 
\begin{equation}
<f({\bf x},{\bf p})>=\int \psi ^{\ast }({\bf x})f({\bf x},\frac{\hbar }{i}%
\nabla )\psi ({\bf x})d^{n}x  \label{expfxp}
\end{equation}
as per standard quantum theory (where in general an operator ordering must
be specified for the expectation value to be well defined).

Finally, we point out another approach that is also natural within this
framework. \ Since the equations of motion in the variables $\psi $ and $%
\psi ^{\ast }$ are linear, it is natural to investigate the group of
canonical transformations that preserve the linearity of these equations. \
This leads to considering the group of transformations of the form 
\begin{eqnarray*}
\rho ({\bf y}) &=&\int K({\bf x},{\bf y})\psi ({\bf x})d^{n}x \\
\rho ^{\ast }({\bf y}) &=&\int K^{\ast }({\bf x},{\bf y})\psi ^{\ast }({\bf x%
})d^{n}x . 
\end{eqnarray*}
Using (\ref{poissondelta}), a simple calculation leads to the following
condition for the transformation to be canonical, 
\begin{eqnarray*}
\left\{ \rho ,\rho ^{\ast \prime }\right\} &=&\int \left[ \frac{\delta \rho 
}{\delta \psi ^{\prime \prime }}\frac{\delta \rho ^{\ast \prime }}{\delta
\psi ^{\ast \prime \prime }}-\frac{\delta \rho }{\delta \psi ^{\ast \prime
\prime }}\frac{\delta \rho ^{\ast \prime }}{\delta \psi ^{\prime \prime }}%
\right] d^{n}x^{\prime \prime } \\
&=&\int K({\bf x}^{\prime \prime },{\bf y})K^{\ast }({\bf x}^{\prime
\prime },{\bf y}^{\prime })\, d^{n}x^{\prime \prime }=\delta ({\bf y}-%
{\bf y}^{\prime }).
\end{eqnarray*}
This is the condition for a transformation to be unitary. \ Arguments
similar to the ones discussed above can then be used to single out the
choice 
\[
K({\bf x},{\bf y})=\frac{1}{\left( 2\pi \hbar \right) ^{-n/2}}\exp (i{\bf x}%
\cdot {\bf y}/\hbar ) 
\]
which corresponds to the transformation that leads to the momentum space
representation. \ 

\section{Conclusions}

We have shown that an exact uncertainty principle, formulated in the form
that the strength of the momentum fluctuations is inversely correlated with
the uncertainty in position, leads from the classical equations of motion to
the Schr\"{o}dinger equation. \ The assumptions that we used for this fall
into three main categories: maximal randomness [Eqs. (\ref{ranassump})
and (\ref{systemindep})]; an exact uncertainty principle [Eq. (\ref{deltaN})]; 
and causality [Eq. (\ref{intPf})].  An alternative derivation is given
in \cite{bam}.

The additional term in the Lagrangian is essentially the Fisher information,
originally introduced by Fisher \cite{Fisher25} as a measure of ``intrinsic
accuracy'' in statistical estimation theory. \ This Fisher information term
was derived using an information theoretic approach in \cite{Reginatto98}.
 \ The connection between Fisher information and quantum
mechanics has been developed further in \cite{Hall00}, where it is shown
that the Fisher information is proportional to the difference of a classical
and quantum variance (thus providing a measure of nonclassicality), and to
the rate of entropy increase under Gaussian diffusion (thus providing a
measure of robustness). The operator formulation of the exact uncertainty
relation in Eq. (\ref{eurel}) is studied in detail in \cite{hallpre},
including its extension to entangled systems.
\ We point out that in all of these references the
Fisher information is defined in the usual way, that is, as a functional of
the probability distribution -- and therefore, one should not confuse it
with the quantity by the same name that appears in Frieden \cite{Frieden99},
which is essentially a generalized Fisher information defined for
wavefunctions and proportional to the quantum kinetic energy.

It is worth noting that the approach here, based on exact
uncertainty, is rather different from other approaches which assign
physical meaning to fields $P$
and $S$ related to the wavefunction.
For example, in the de Broglie-Bohm approach \cite{BB}, there are {\it
no}
momentum fluctuations, and the classical equations of motion for $P$ and
$S$ are instead modified by adding a mass-dependent ``quantum
potential'',
$Q$, to the classical potential term in the Hamilton-Jacobi equation.
The form of this quantum potential is left unexplained, and is
interpreted as
arising from the influence of an associated wave acting on the system.
Similarly, while Bohm and Vigier generalise Bohm's original
formalism to permit fluctuations of momentum about $\nabla S$, this is
merely to ensure that an ensemble of such particles will quickly evolve to
have a stable distribution given by the modulus-squared of the associated
wave \cite{BomVig54}.  
In contrast, in the exact uncertainty approach $\nabla S$ is an
{\it average} momentum, the form
of an additional kinetic energy term arising from random momentum
fluctuations is {\it derived}, and no associated wave is assumed.  The
formal connection between the two approaches is the relation
\[ \delta(L-L_C) = \int dt\,d^nx\,\,Q \,\delta P.\]

The exact uncertainty approach is also very different from the
stochastic mechanics approach \cite{NE}.  The latter postulates the
existence of a classical stochastic process in configuration space, with
a drift
velocity assumed to be the gradient of some scalar, and defines
an associated time-symmetric 
``mean acceleration'' ${\bf a}$ in terms of averages over both
the stochastic process
and a corresponding time-reversed process, which is postulated
to obey Newton's law
$m{\bf a}=-\nabla V$. In contrast, the exact uncertainty 
approach does not rely on a
classical model of fluctuations, nor on a new definition of
acceleration,
nor on properties of stochastic processes running backwards in time. 
Indeed, as remarked in section 3.1,
one may view the introduction of ${\bf N}$ as a means of
effectively eliminating the notion of trajectories, differentiable or
otherwise, from the classical
hydrodynamical formulation of section 2.
The formal connections between the approaches are
\[ \nabla S=m{\bf u}, \hspace{1.5cm}(\Delta N)^2 = m^2\langle{\bf v\cdot
v}
\rangle,\]
where ${\bf u}+{\bf v}$ and ${\bf u}-{\bf v}$ are the drift
velocities of the forward-in-time and backward-in-time processes
respectively.
It should be noted that $\langle {\bf u\cdot v}\rangle\neq 0$, and
hence, noting Eq. (5), one cannot identify $m{\bf v}$ with the random momentum
fluctuation
${\bf N}$.

In \cite{Reginatto98} it was suggested that the Fisher information term
represented an ``epistemological'' contribution to the action, which in the
context of the present analysis can be interpreted as reflecting a lack of
detailed knowledge of nonclassical momentum fluctuations. \ In our approach
we do not attempt to provide a ``realistic'' model of such fluctuations,
which would at any rate require a whole new (and nonlocal) 
theory that goes beyond quantum
mechanics. \ Our approach to understanding quantum mechanics is therefore
different from other descriptions based on the postulate of an underlying
stochastic process, such as stochastic mechanics \cite{NE}. \ What our
analysis primarily offers is a new way of viewing the uncertainty principle
as {\it the} key concept in quantum mechanics. \ While it is true that no
one before quantum mechanics would think of taking an uncertainty principle
as a fundamental principle, our analysis is valuable in that it enforces the
importance of the uncertainty principle in distinguishing quantum mechanics
from classical mechanics -- in a sense, it says that the uncertainty
principle is {\it the} fundamental element that is needed for the transition
to quantum mechanics.

\end{document}